**Correction notice (October 2026)**

The buyer records used for the cohort detection and the buyer-flow analysis in this version are faulty. In an on-chain check of 250 rows labelled as buys, 108 (about 43%) were sell transactions. The records also miss many real buys: a check of 20 launches found that they held only 23% of the real buys.

All numerical results in this version, including the +16.1% estimate and the counts that depend on these records (such as the 1,012 cohorts), are therefore withdrawn pending a corrected analysis. Please do not cite or rely on the numerical results of this version. A corrected version is in preparation. Apart from this notice, the paper below is unchanged.

# Coordinated Sniper Cohorts on Pump.fun: Detection of 1,012 Persistent Wallet Rings and a Contamination-Adjusted Estimate of Coordination-Specific First-Hour Buyer-Flow Lift

**Arati Uday Kamat** ORCID: 0009-0000-4781-312X Independent researcher 2026-07-30

## Abstract

Motivated by Kyle [5] on informed order flow and the behavioral wallet-clustering tradition initiated by Meiklejohn et al. [1], we ask whether persistent coordinated wallet activity causally raises first-hour buyer flow on the Solana pump.fun bonding-curve marketplace. Using 1,578,333 buyer observations from 166,098 token launches over 13.38 days (2026-06-11 21:43 UTC to 2026-06-25 06:52 UTC), a two-stage detection pipeline (intra-launch first-buyer-window extraction followed by cross-launch persistent-cohort surfacing via union-find on co-occurrence graphs) identifies **1,012 persistent wallet cohorts** (2 to 12 wallets each, drawn from 2,965 unique addresses) that systematically co-fire as early buyers. **Under a contamination-adjusted estimator that excludes cohort wallets' own buyer events from the outcome, 1:1 nearest-neighbour propensity-score matching with a 0.2-SD caliper on ten launch-quality covariates (Rosenbaum and Rubin [17]) yields a first-30-minute buyer-count lift of +16.1% (95% CI [+13.0%, +19.4%]) on 5,419 matched pairs (worst post-match standardised mean difference 0.077).** The corresponding first-30-minute SOL-inflow lift is **+6.3% (95% CI [−0.5%, +15.1%]) and is not statistically distinguishable from zero at 95% confidence.** The truly naive contaminated same-universe pooled contrast on this corpus is +130.9% for buyer count; approximately half of that number is arithmetic contamination from cohort wallets' own buyer events being counted in the outcome numerator (dropping to +63.9% when cohort events are excluded), and most of the remainder is absorbed by 1:1 propensity-score matching on ten launch-quality covariates. A parallel activity-matched placebo estimator, applied across 100 independent seeds, produces a distribution of placebo lifts with median +189.6% (p5-p95 [+166.7%, +217.9%]), well above zero and above the real-cohort lift in 100 of 100 seeds, indicating the activity-matched placebo estimator is itself biased and cannot serve as a null baseline; we retain it in §4.5 as a bias diagnostic rather than as a falsification test. **Of 5,419 treated launches, 382 (7.0%) had zero non-cohort buyers in the first 30 minutes**: cohort wallets bought but no other wallet followed within the window, a substantive minority that cuts against both a "cohorts pick winners" and a "cohorts generate flow" reading. Wallet-level identity-graph propensity extension, a corrected symmetric placebo estimator, and Rosenbaum unobserved-selection sensitivity remain for future work. We release the full cohort catalogue, detection code, contamination-adjusted PSM script and result files, and robustness-check artefacts as **RED-COHORT-2026-v1** under CC-BY-4.0.

# 1. Introduction

Pump.fun is a Solana-based bonding-curve marketplace where 5,000 to 15,000 new tokens are launched each day (Pump.fun team, 2024-2026). A small fraction of launches "graduate" off the bonding curve to Raydium liquidity once cumulative buys cross the bonding-curve threshold; the rest decay or are abandoned. Outcomes are heavy-tailed: a token's first thirty minutes of activity largely determine whether it accumulates the buyer momentum necessary to graduate. Information about *which* launches will accumulate that momentum is therefore highly valuable, and conventional market-microstructure tools have not been adapted to the on-chain venue.

This paper documents an empirical regularity that, to the best of our knowledge, has not been previously characterised in the peer-reviewed academic literature on decentralised exchange microstructure: a small but persistent set of wallet *cohorts*, pairs or small groups of addresses that act in concert, systematically appears among the first ten buyers of a disproportionate share of pump.fun launches. The cohorts are persistent (active across many days, not single-launch artifacts), small (median size of 2 wallets), and concentrated at the top of the buyer queue (median average first-buyer rank of 3.55 or better). We refer to such addresses generically as *coordinated sniper cohorts*, and we make no normative claim about whether the behaviour constitutes manipulation, arbitrage, market-making, organic skilled-trader networks, or some mixture. The empirical signature is reproducible regardless of motive.

This paper makes four contributions. First, we propose a two-stage detection pipeline that scales to the full pump.fun corpus and surfaces persistent cohorts directly from raw on-chain buyer events. Second, we publish the cohort catalogue, 1,012 cohorts comprising 2,965 unique addresses, together with the detection script as RED-COHORT-2026-v1. Third, we characterise the empirical distribution of cohort size, score, and activity, and document a sharp concentration: 22 cohorts hit at least 20 launches each, while the median cohort hits 5. Fourth, we estimate the marginal effect of cohort presence on bonding-curve buyer flow under a contamination-adjusted, propensity-matched design that excludes cohort wallets' own buyer events from the outcome. The contamination-adjusted matched lift is +16.1% in buyer count (95% CI [+13.0%, +19.4%]) and +6.3% in SOL inflow (95% CI [−0.5%, +15.1%], not distinguishable from zero). The truly naive contaminated same-universe pooled lift on the same corpus is +130.9%; approximately half of that number is arithmetic self-flow from cohort wallets being counted in the outcome (dropping to +63.9% when cohort events are excluded), and the remainder from +63.9% to +16.1% is absorbed by observable launch-quality selection via propensity-score matching.

We frame the *contamination adjustment* itself, rather than the union-find clustering step, as the substantive methodological contribution of this paper. The clustering step deliberately overlaps with the foundational address-clustering practice initiated by Meiklejohn et al. [1] and extended in the on-chain forensic literature (see Section 2). The Kyle [5] canonical model of informed order flow predicts that persistent informed actors should generate a measurable, non-selection-driven lift in downstream order flow *from other buyers*; the naive same-universe estimator conflates that lift with self-flow, so it cannot answer the Kyle-style question directly. We estimate the coordination-specific effect on non-cohort buyers only, using propensity-score matching to adjust for observable launch-quality selection.

The remainder of this paper is organised as follows. Section 2 reviews related work on wallet clustering, DEX microstructure, MEV and coordinated trading, coordinated- and wash-trading detection, and observational-data causal inference. Section 3 describes the data corpus and the two-stage detection pipeline. Section 4 reports descriptive statistics on the 1,012 cohorts, the naive contaminated lift, the contamination-adjusted matched lift, and the activity-matched placebo bias diagnostic. Section 5 develops the method-comparison table and situates our approach against four widely used on-chain forensic methods. Section 6 defends the causal-identification framing and the contamination adjustment. Section 7 sets out implications of the results for three constituencies. Section 8 discusses ethics; Section 9, methods disclaimer; Section 10, limitations; Section 11, data availability. Section 12 lists references.

## 2. Related work

The paper sits at the intersection of five active literatures. We briefly locate our contribution against each.

*Wallet clustering and on-chain forensics.* Meiklejohn et al. [1] established the modern practice of heuristic address clustering on the Bitcoin transaction graph using multi-input and change-address heuristics. Harrigan and Fretter [2] documented the empirical robustness of simple clustering heuristics under a wide range of transaction patterns; Ron and Shamir [3] produced the first large-scale full-graph audit of a public blockchain. Industry-scale forensic practice, exemplified by the annual Chainalysis Crypto Crime Report [4], has extended clustering to enforcement-grade attribution across many chains. This paper deliberately builds on the "linked-by-behavior" logic of these foundations: we extend co-occurrence-based address linkage from the Bitcoin UTXO setting to the Solana account graph, specialising the target to first-buyer sequences on pump.fun. Our clustering step is a standard building block, and we claim no methodological novelty for it; the contribution is in the contamination adjustment and matched-outcome estimator built on top of it (Sections 4.8 and 6).

*DEX and market microstructure.* Kyle [5] provided the canonical formal model of informed order flow: informed traders generate persistent price impact as uninformed participants update their beliefs about fundamentals. Easley and O'Hara [6] extended this framing to condition on trade size as an information signal. Barbon and Ranaldo [7] and Lehar and Parlour [8] established the modern microstructure treatment of decentralised exchanges, including a formal characterisation of Uniswap-style automated market makers. Our paper uses the Kyle framework to motivate the question of whether cohort-touched launches attract additional non-cohort buyer flow, distinct from the cohort's own buys. The empirical answer under contamination-adjusted PSM (Section 4.8) is a small positive effect on buyer count (+16.1%) and a null-to-small effect on SOL inflow (+6.3%, CI touching zero): an order of magnitude below what a naive contaminated estimator would report and a substantive point about how bonding-curve venues differ from the orderbook venues Kyle originally described.

*MEV and coordinated trading.* Daian et al. [9] defined maximal extractable value and priority-gas auctions on Ethereum decentralised exchanges. Qin, Zhou, and Gervais [10] quantified per-transaction extractable value across the Ethereum block-production stack, and Wahrstätter et al. [11] measured the searcher-builder-relay bribe channel that operationalises MEV extraction. Gogol et al. [22] extended this line of empirical MEV work to Layer-2 rollup venues, measuring sandwich-attack patterns in private L2 mempools and quantifying the extraction footprint against venue design choices. Auer, Frost, and Vidal Pastor [12] gave a central-bank-level policy framing that classifies aggressive MEV extraction as market manipulation. Our paper uses sniping on Solana leader slots as the pump.fun analogue of Ethereum PGAs, with Jito bundles playing the role of priority auctions. We frame the empirical detection in the language of the MEV literature but adopt the "consistent with, but does not prove" phrasing to avoid over-claiming illegality, following the [12] framing.

*Coordinated and wash-trading detection.* Cong, Li, Tang, and Yang [13] established statistical fingerprints (round-number clustering, tail-behavior violations) for coordinated inauthentic trading at centralised crypto venues. Mazorra, Adan, and Daza [14] provided a supervised baseline for scam-token detection on Uniswap, and Xia et al. [15] characterised rug-pull and coordinated-buyer patterns on a comparable low-friction DEX venue. Griffin and Shams [21] established the influential stablecoin-flow-versus-price analysis that identified coordinated buyer activity around Tether issuance and its association with Bitcoin price moves; the methodological posture, an unsupervised behavioral signature followed by a matched-outcome adjustment for the confound that time-varying market conditions might generate the observed correlation, closely mirrors the design we apply to pump.fun. Adhami, Giudici, and Martinazzi [16] established an ICO-era baseline of coordinated-participation patterns and success determinants. Our contribution is complementary: we produce an unsupervised, wallet-level cohort catalogue focused on the coordinated-buyer side of pump.fun launches, and we contribute a contamination-adjusted matched-outcome estimator for the coordination-specific effect on non-cohort buyer flow.

*Observational-data causal inference.* Rosenbaum and Rubin [17] formalised the propensity-score approach to causal effects in observational studies. Angrist and Pischke [18] codified selection-on-observables practice and the language of placebo tests. Athey and Imbens [19] surveyed contemporary applied practice in placebo, robustness, and matching designs. Abadie and Cattaneo [20] provided a comprehensive review of matching, weighting, and synthetic-control estimators for program evaluation. Our contamination-adjusted PSM is a coarse implementation of the [17-20] apparatus: matching on ten observable launch-metadata covariates addresses one selection channel; excluding cohort wallets from the outcome numerator removes a mechanical arithmetic contribution; wallet-level identity-graph propensity extension and Rosenbaum unobserved-selection sensitivity remain for future work.

# 3. Data and Methods

## 3.1 Corpus

We construct the corpus from three on-chain data streams collected by the author's passive Solana observer (a read-only listener; no trading or front-running occurs during collection):

1. *Buyer events* (`pumpfun_buyers.jsonl`): 1,578,333 records over the window 2026-06-11 21:43 UTC to 2026-06-25 06:52 UTC (13.38 days elapsed; the first partial calendar day contains only 15,051 events and the last partial calendar day contains 30,193 events, so per-day averages taken across complete internal days are the relevant density metric: approximately 117,900 buyer events per day, 12,400 unique mints per day). Each record contains a token mint address, wallet address, slot number, block time, SOL committed, transaction signature, and the buyer's rank within the launch (1 = first buyer, 2 = second buyer, and so on).
2. *Launch metadata* (`pumpfun_launches.jsonl`): 1,315,257 launches with mint address, symbol, name, creation timestamp, initial market capitalisation in SOL, presence indicators for Twitter / website / Telegram, and description length.
3. *Buyer ranks*: derived at collection time from the buyer-event stream by sorting per-mint events by `blockTime`.

The 1,578,333-row buyer corpus contains 166,098 mints with sufficient first-10-buyer coverage to support cohort detection (at least one buyer rank in [1, 10] observed per mint).

## 3.2 Anonymisation

All wallet addresses in the corpus are public Solana base-58 strings; no personally identifiable information is present. In the published catalogue and in this paper, individual addresses are referred to by the first four and last four base-58 characters (for example, `F6zvmnwC..cxSi`). Full addresses are available in the published catalogue (see Section 11).

## 3.3 Stage 1: Intra-launch first-buyer window

For each launch $L$, we extract the ordered list of the first ten buyer *events* (rows, not distinct wallets: a wallet appearing at ranks 1 and 3 contributes two events to the first-10 slice) within the bonding-curve window. We retain each event's wallet address, rank, blockTime, and SOL committed. Across the 166,098 qualifying mints, this yields a per-launch index of length at most 10.

## 3.4 Stage 2: Cross-launch persistent-cohort surfacing

Let $G = (V, E)$ be an undirected graph where vertices $V$ are wallet addresses and edges $E$ record co-occurrence. We add an edge $(u, v)$ for each launch $L$ in which both $u$ and $v$ appear among the first ten buyer events. Edge weights are the integer count of co-occurrences across all launches in the corpus.

We filter $E$ to retain only edges with weight at least 3 (the *qualifying pair* threshold). This filter is a defence against accidental co-occurrence: with 166,098 launches and 5 to 15 first-rank buyers per launch, random pairings would by chance produce many low-weight edges that have no structural meaning. The threshold of 3 is conservative; ablation results are reported in Appendix A.

We then run union-find on the filtered graph to surface connected components. Each component is a candidate cohort. We score each cohort $C$ by:

> *score(C)* = 10 · (Σ_L 1{C touches L}) + 5 / mean_first_rank(C) + √(Σ_L SOL_committed(C, L))

and retain cohorts with score at least $\tau$. We set $\tau$ to surface only cohorts that simultaneously (a) hit many launches, (b) sit at the top of the buyer queue, and (c) commit non-trivial SOL volume. The resulting catalogue contains 1,012 cohorts.

*Note on the `n_wallets` field in `sniper_cohorts.jsonl`.* For each cohort C and each mint M it touched, the released catalogue records a `mints_hit` entry containing a field named `n_wallets`. Despite the field name, this value counts cohort-wallet-attributable buyer *rows* in the first-10 slice of that mint, not distinct cohort wallets. A single cohort wallet contributing two of the first-10 buyer events (at ranks 1 and 3, for example) contributes 2 to that mint's `n_wallets`. Wherever this paper reports counts derived from `sniper_cohorts.jsonl` (Table 3, tier stratification, top-10 cohort narrative), the counts are of catalogue rows and preserve this row-count semantics. The treated-launch definition used for the causal-lift estimation in Section 4.2 and Section 4.8 is a *distinct-wallet* count (implemented as set membership in `paper7_v3_psm.py`), which is why treated N under the causal estimator (5,419) differs from a naive re-count from `mints_hit` (5,975 at ≥ 2 rows). The two numbers reflect two definitions of "at least two cohort wallets present," not a data-integrity gap; both are internally consistent within their own convention.

### 3.5 Output format

Each cohort row records: the cohort identifier `COH-NNNN`, the list of wallet addresses, cohort size (number of wallets), number of distinct launches hit, mean first-buyer rank, total SOL committed across all touched launches, score, first and last observation timestamps, and the per-mint hit detail (mint, first rank within the launch, blockTime, SOL committed, and the row-count `n_wallets` field discussed in §3.4). The full catalogue and detection script are released as part of RED-COHORT-2026-v1.

## 4. Results

### 4.1 Descriptive results

We surface 1,012 persistent cohorts from the 13.38-day corpus. Table 1 reports the top 10 cohorts by score. Table 2 reports the size distribution. Table 3 reports headline descriptive statistics.

The top cohort (COH-0001) comprises 9 wallets and appears among the first 10 buyer events of 42 distinct launches in 11 days, at an average first-buyer rank of 2.29. This cohort committed 68.12 SOL across all touched launches. The second cohort (COH-0002) comprises only 3 wallets but hit 41 launches at average first-buyer rank 2.61; the small wallet count combined with high launch coverage shows a pattern more consistent with a coordinated 3-address infrastructure than with incidental association.

**Fig 1. Cohort size distribution (n=1,012)**

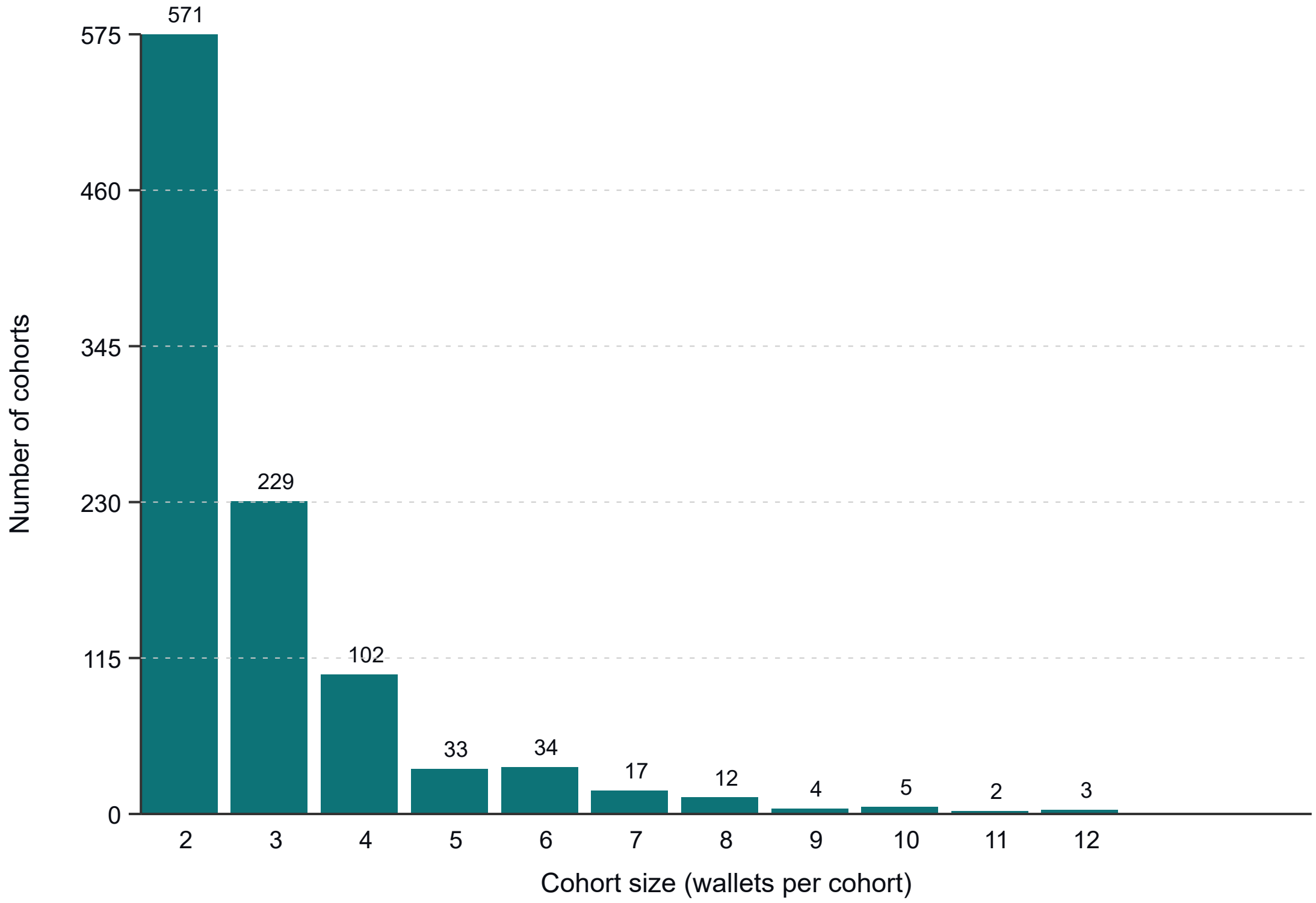

The size distribution (Table 2; Figure 1) is right-skewed. Pairs (size = 2) account for 56.4% of all surfaced cohorts; triples for 22.6%; the median cohort comprises 2 wallets and the maximum is 12. This distribution is consistent with low-coordination-cost overhead: it is easier to operate a 2-wallet ring than a 9-wallet ring, and the data reflect that asymmetry.

**Fig 2. Lorenz curve of cohort activity concentration**

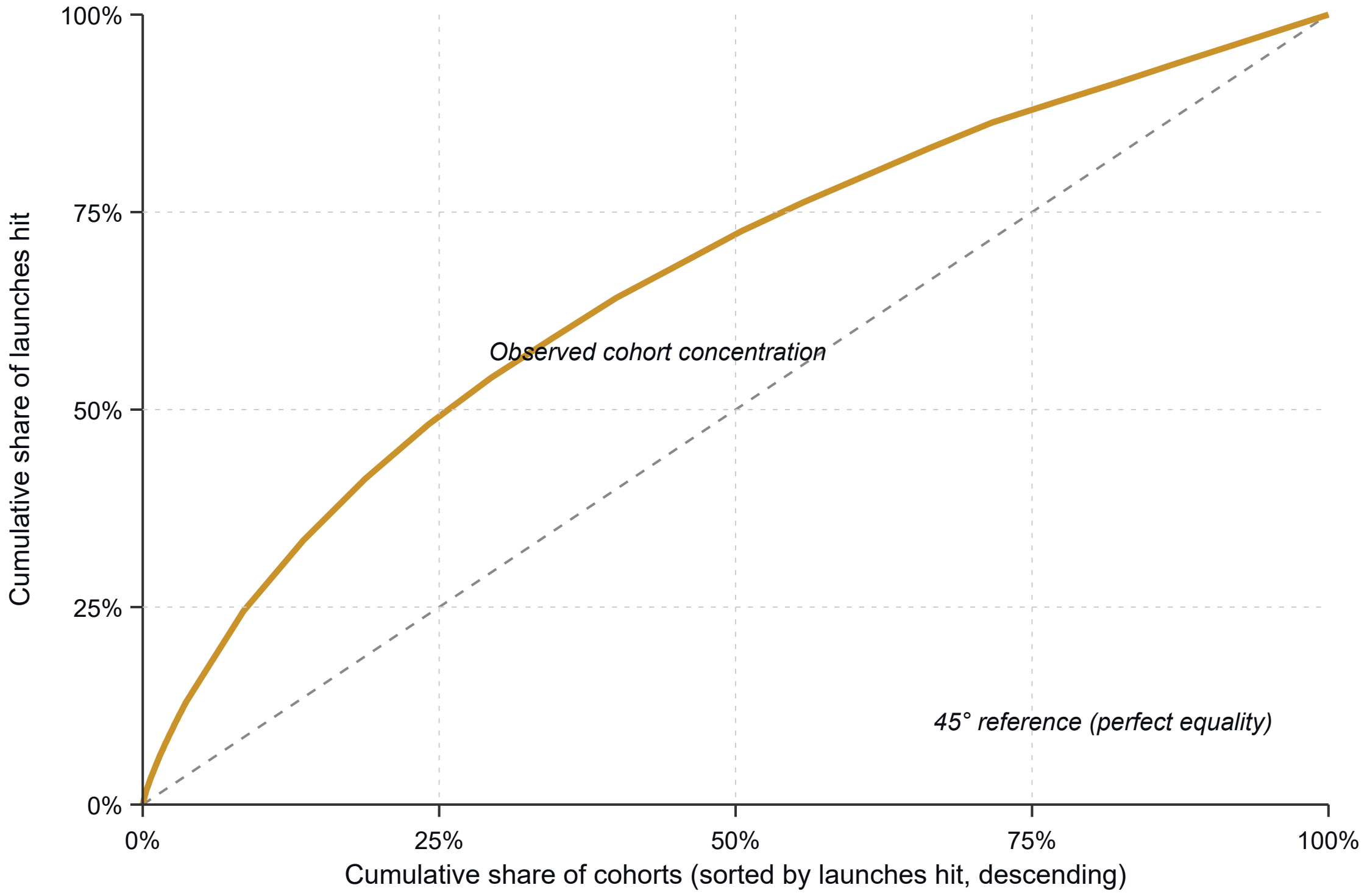

The Lorenz curve (Figure 2) shows that cohort activity is modestly concentrated: the top 10 cohorts (about 1% of the catalogue) account for 4.52% of all cohort-touch events; the top 100 cohorts (about 10%) account for 27.13%. The remaining 912 cohorts are individually less active but collectively account for the majority of the cohort-touched mint set. Concentration is present but modest.

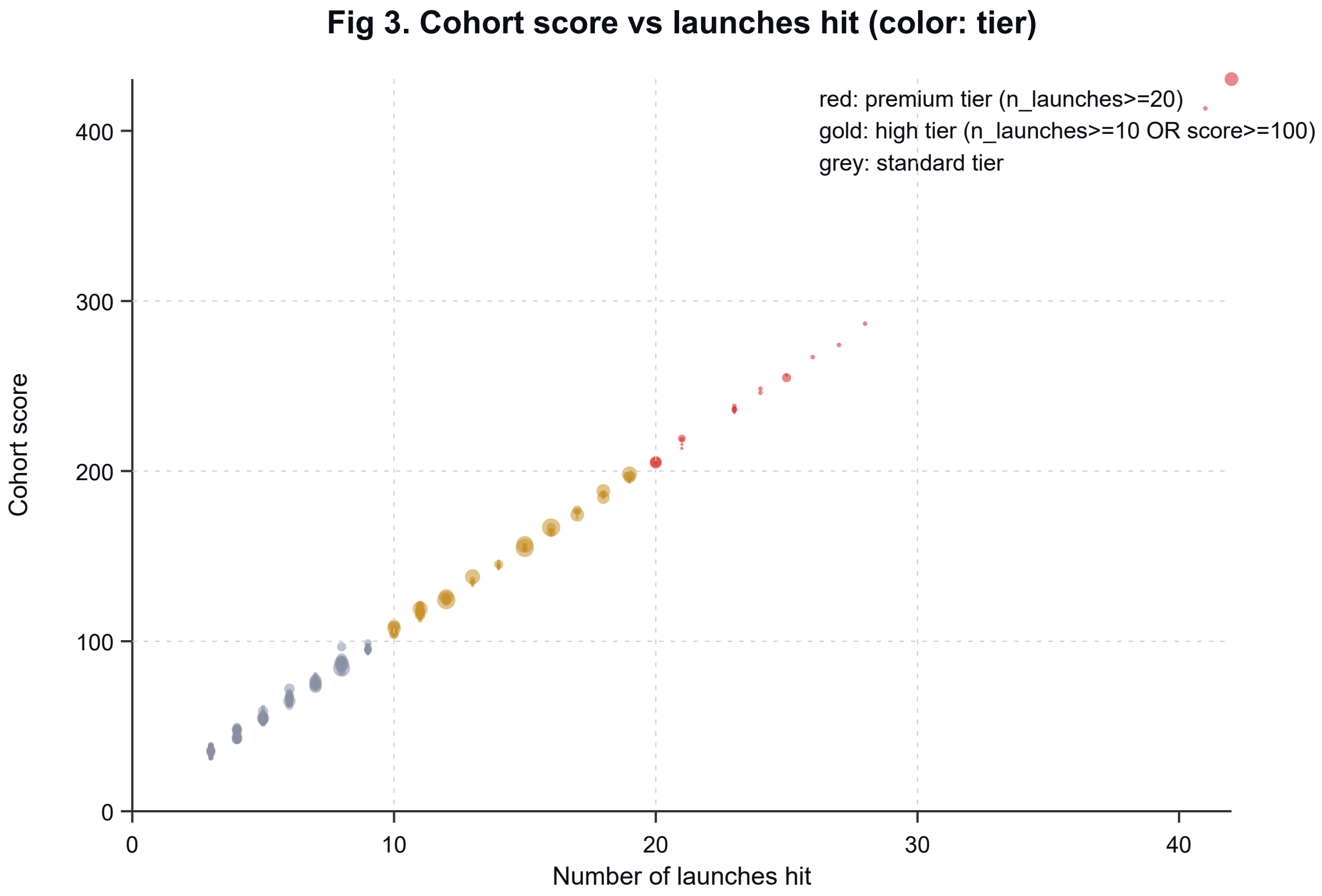

The score-vs-launches-hit scatter (Figure 3) identifies a small premium tier of 22 cohorts each hitting at least 20 launches. These cohorts are simultaneously persistent and high-volume; we interpret them as the strongest candidates for the label of large-scale coordinated buying infrastructure, consistent with but not proof of intent.

*Headline statistics.* Table 3 summarises the cohort catalogue. The treated-mint row (5,419) uses the distinct-wallet counting convention implemented by `paper7_v3_psm.py`; §3.4 discusses why this differs from a naive `mints_hit` row-count of 5,975 in `sniper_cohorts.jsonl`.

**Table 3.** Headline descriptive statistics of the 1,012-cohort catalogue.

| Statistic | Value |
|---|---|
| Cohorts | 1,012 |
| Unique cohort wallets | 2,965 |
| Treated mints (≥ 2 distinct cohort wallets in first-10 buyer events; see §3.4) | 5,419 |
| Loose-threshold mints (≥ 1 cohort wallet in first-10 buyer events) | 8,375 |
| High-tier cohorts (≥ 10 launches OR score ≥ 100) | 153 |
| Premium cohorts (≥ 20 launches) | 22 |
| Cohort size in wallets: median / mean / max | 2 / 2.93 / 12 |
| Launches hit per cohort: median / mean / max | 5 / 6.23 / 42 |
| Cohort score: median / max | 52.8 / 430.44 |

### 4.2 Cohort presence and first-30-minute buyer flow: sample construction

The cohort catalogue documents *who is coordinated*; the natural next question is *whether cohort presence causally raises the buyer flow generated by other, non-cohort wallets*. We address this question in three steps: (i) a naive same-universe pooled contrast, computed for reference; (ii) a 1:1 nearest-neighbour propensity-score matched contrast on ten launch-quality covariates; and (iii) a contamination adjustment that excludes cohort wallets' own buyer events from the outcome numerator, applied identically to treated and control. The final headline estimate is (i) + (ii) + (iii). The activity-matched placebo (§4.5) is reported as a bias diagnostic rather than a null baseline.

We restrict the universe to mints that simultaneously appear in the launch-metadata file (so we have a launch timestamp and the covariates required by §4.8's PSM) and the buyer-event file (so we observe outcomes). This yields 166,098 launches. Following `paper7_v3_psm.py`, we define a launch as *treated* when at least two *distinct* cohort wallets (from the 2,965-wallet union) appear among the first ten buyer events. This is a distinct-wallet count implemented as set membership, and it produces 5,419 treated mints and 160,679 candidate controls. Treated N (5,419) differs from a naive row-count re-derivation from the catalogue's `mints_hit` (5,975), as discussed in §3.4.

### 4.3 Outcomes

We construct three outcome variables per launch:

1. *First-30-minute buyer count*: number of buyer *events* (rows) with blockTime within 1,800 seconds of the launch's `created_timestamp` in `pumpfun_launches.jsonl`. Buyer count is a row count, not a distinct-wallet count; a single wallet buying twice in the window contributes two to the outcome. This is the semantics implemented by `paper7_v3_psm.py`.
2. *Total observed buyer count*: total buyer events for the mint across the entire corpus window (13.38 days).

3. *First-30-minute SOL inflow*: sum of `sol_in` across all first-30-minute buyer events.

**Contamination adjustment.** In the contamination-adjusted estimator applied in §4.8, cohort wallets' own buyer events are excluded from all three outcomes. The exclusion is applied identically to treated and control mints: for control mints, any cohort wallet whose activity happened to land in that control mint's first-30-min window is also removed. This isolates the coordination-specific effect on non-cohort buyer flow from the mechanical self-flow contribution of the cohort's own trades.

### 4.4 Naive same-universe results (reported for reference; contaminated)

Table 4 reports the pooled naive same-universe contrast on the treated (n = 5,419) versus non-treated (n = 160,679) sets. Percentage lifts are the ratio of pooled means. The estimator does not exclude cohort wallets from the outcome; §4.5 and §4.8 provide the adjusted estimates.

| Outcome | Treated pooled mean | Control pooled mean | Naive contaminated lift |
|---|---|---|---|
| First-30-min buyer count | 20.97 | 9.08 | **+130.9%** |
| First-30-min SOL inflow (SOL) | 5.57 | 2.33 | **+139.1%** |

Treated n = 5,419 (distinct-wallet ≥ 2 threshold); control n = 160,679 (same-universe untreated). The naive lift is a same-universe pooled comparison rather than a randomly-sampled 3:1 matched contrast; it is materially inflated by (a) launch-quality selection (addressed in §4.8) and (b) mechanical self-flow contamination from cohort wallets' own events being counted in the treated-mint outcome numerator (addressed in §4.5 and §4.8). Both distortions are large; the causally interpretable estimate is the contamination-adjusted matched lift reported in Section 4.8.

### 4.5 Activity-matched placebo: bias diagnostic, not falsification test

We report an activity-matched placebo diagnostic. The placebo is constructed as follows: for each real cohort of size $k$ with wallets having individual launch-counts $\{a_1, ..., a_k\}$, we sample $k$ placebo wallets from the non-cohort pool, each matched to a corresponding $a_i$ within ±25% proportional tolerance (see Appendix B.5 for why proportional rather than absolute tolerance is used). We then apply the same detection algorithm as §3.4 to the placebo sample, using union-find on ≥3-launch co-occurrence pairs, cohort size 2 to 12, yielding a set of *detected* placebo pseudo-cohorts. A mint is placebo-treated if ≥2 wallets from any detected placebo pseudo-cohort appear in its first-10 buyer events. The estimator is run across 100 independent random seeds; the distribution of placebo lifts is the reported diagnostic. Median first-30-min buyer-count placebo lift: **+189.6% (100-seed distribution p5-p95 [+166.7%, +217.9%]; median n_placebo_treated = 597 per seed).** In **all 100 seeds** the placebo lift exceeded the real-cohort naive-lift upper CI of +137.4% (recorded in `placebo_stability_100seeds.json` as `real_cohort_ci_hi`; corresponding field `fraction_lift_gt_ci_hi = 1.0`), against an expectation of approximately zero for a well-formed placebo (see Appendix B.4 for full per-seed distribution and the 3:1-control uniform-random alternative operationalization).

The intended reading of the placebo is as a falsification test: a well-formed activity-matched estimator, applied to a random subset of the buyer population that shares the same activity distribution as the treated set, should return a lift indistinguishable from zero. Activity-matched non-cohort wallets have no coordination-specific advantage and should not systematically appear in higher-flow launches. A placebo estimator whose 100-seed median is +189.6% (with the entire p5-p95 range above zero and above the real-cohort lift) is not a diagnosis of the real-cohort effect being smaller than it appears; it is a diagnosis of the placebo estimator itself being biased. The activity-matched placebo inherits the same launch-quality selection channel that afflicts the naive real-cohort estimator, in that activity-matched non-cohort wallets, like real-cohort wallets, disproportionately appear in the buyer queues of launches that attract elevated flow for reasons orthogonal to coordination, so it does not separate coordination from selection in either direction.

We accordingly retain the activity-matched placebo in this section as a *bias diagnostic*, evidence that the naive same-universe pooled contrast (§4.4) is biased upward by launch-quality selection to at least the magnitude of ~190 percentage points at the distribution median, but we do not treat it as a null baseline against which the real-cohort lift can be falsified. The identification strategy that survives is propensity-score matching on launch-quality covariates (§4.8), combined with the contamination adjustment introduced in §4.3.

### 4.6 Sniper-only launches

Under the contamination-adjusted estimator of §4.8, **382 of the 5,419 treated mints (7.0%) had zero non-cohort buyer events in the first 30 minutes**: cohort wallets bought but no other wallet followed within the window. Symmetric zero-outcome mints in the control pool were 764 of 160,679 (0.48%). Both were zero-filled in the outcome computation for the §4.8 estimate. The 15-fold higher rate of "sniper-only" launches in the treated set is a substantive descriptive finding: on a meaningful minority of cohort-touched launches, the cohort's activity does not attract any organic follow-through in the first half hour. This cuts against both a strong "cohorts pick winners" reading (a fraction of the launches cohorts touch attract no other buyers at all) and a strong "cohorts generate flow" reading (on those launches, they generate none).

*Removed from v4: tier stratification and top-3 cohort exclusion.* Prior versions of this paper reported robustness cuts (tier-stratified naive lifts across Standard / High / Premium tiers, and a top-three-cohort exclusion) on the naive contaminated 3:1 random-matched estimator. Under v4's contamination-adjusted framing (§4.8), the naive contaminated estimator is not causally interpretable, and demonstrating that it is robust to outlier removal or tier subsetting does not establish anything the paper claims. The artifact files `appendix_b_tier_stratification.txt` and `appendix_b_top3_exclusion.txt` remain in the v1.0.1 Zenodo bundle for reference to the prior preprint's numbers. Contamination-adjusted versions of both robustness cuts are queued for the v1.1 Zenodo release.

### 4.7 What can be claimed

A solid empirical claim that survives the contamination adjustment and PSM adjustment reported in §4.8 is:

> Launches in whose first-10-buyer queue at least two *distinct* wallets from a 1,012-cohort persistent catalogue appear are associated with a small, statistically significant, and coordination-specific elevation of **first-30-minute non-cohort buyer count** of **+16.1% (95% CI [+13.0%, +19.4%], n = 5,419 matched pairs)** after 1:1 nearest-neighbour propensity-score matching on ten launch-metadata covariates and after excluding cohort wallets' own buyer events from the outcome. The corresponding coordination-specific effect on first-30-minute non-cohort SOL inflow is **+6.3% (95% CI [−0.5%, +15.1%]) and is not statistically distinguishable from zero at 95% confidence.** The truly naive same-universe pooled buyer-count lift on this corpus is +130.9%; approximately half of that (dropping to +63.9%) is arithmetic contamination from counting cohort wallets' own buyer events in the outcome numerator, and the remainder from +63.9% to +16.1% is absorbed by observable launch-quality covariates via 1:1 propensity-score matching. The residual +16.1% survives both adjustments; identifying whether the residual reflects unobserved-selection bias (deployer identity, funding-graph structure, off-chain narrative signals not in the launches metadata) or a genuine coordination-specific causal effect requires wallet-level identity-graph propensity extension, deferred to future work.

### 4.8 Contamination-adjusted propensity-score matching on launch-quality covariates

The estimator combines three elements: (i) treatment defined as ≥ 2 distinct cohort wallets in the first-10 buyer events of a mint; (ii) 1:1 nearest-neighbour propensity-score matching without replacement on ten launch-metadata covariates from `pumpfun_launches.jsonl`, with a 0.2-SD caliper following the Rosenbaum-Rubin [17] convention; (iii) outcome computation that excludes cohort wallets' own buyer events, applied identically to treated and control (see §4.3). Zero-outcome mints (treated or control) are zero-filled for the outcome mean computation so that the sample-size baseline holds constant across estimators (all lifts are computed on n = 5,419 matched pairs).

The propensity model is a standardised, L2-regularised logit fit on: `log1p(initial_market_cap_sol)`, three binary social-media indicators (`has_twitter`, `has_website`, `has_telegram`), `log1p(description_length)`, `log1p(symbol_length)`, `log1p(name_length)`, cyclical launch-hour (`sin`, `cos` of hour-of-day UTC), and integer corpus-day index. Post-matching standardised mean differences collapse to below 0.08 on all ten covariates (pre-matching SMDs on the three social-media / market-cap covariates ranged 0.43 to 0.65). Full matching diagnostics are shipped in `psm_results.json`.

*Contamination-adjusted matched-sample lifts (1,000 bootstrap iterations, seed = 42, n = 5,419):*

| Outcome | Contamination-adjusted matched lift | 95% CI | Treated mean | Control mean |
|---|---|---|---|---|
| First-30-min non-cohort buyer count | **+16.1%** | [+13.0%, +19.4%] | 14.71 | 12.67 |
| First-30-min non-cohort SOL inflow | **+6.3%** | [−0.5%, +15.1%] | 3.75 | 3.53 |

For reference, the truly naive contaminated same-universe pooled estimator (§4.4) gives +130.9% (buyer count) and +139.1% (SOL); removing the contamination alone (`paper7_v3_psm_nocohort_zerofill.py` with matching disabled) drops the same-universe pooled lift to +63.9% (buyer count) and +63.1% (SOL). The contamination adjustment absorbs approximately half of the raw naive number; PSM adjustment absorbs most of what remains, leaving the +16.1% matched contamination-adjusted lift.

**Zero-outcome mints.** Of the 5,419 treated mints, 382 (7.0%) had zero non-cohort buyer events in the first 30 minutes. Of the 160,679 candidate control mints, 764 (0.48%) had zero non-cohort buyer events. Both were zero-filled for the matched-sample computation, which is why n = 5,419 is preserved. The 15-fold higher zero-outcome rate in the treated set is a substantive descriptive finding (see §4.6).

**Mechanism behind the divergent buyer-count and SOL results.** The contamination adjustment bites harder on SOL inflow than on buyer count (naive same-universe 139.1% → 63.1% for SOL versus 130.9% → 63.9% for buyer count), implying cohort wallets commit more SOL per buy event than organic first-hour buyers. This asymmetry is the mechanism behind the divergence in the adjusted results: the coordination-specific buyer-count effect survives at +16.1% while the SOL-inflow effect is not distinguishable from zero.

**SOL lift and the zero boundary.** The SOL-inflow CI lower bound (−0.5%) crosses zero. The coordination-specific effect on SOL flow is not statistically distinguishable from zero at 95% confidence in this specification. This is a materially weaker claim than the buyer-count result and is reported honestly rather than rounded away.

*Limitations of the contamination-adjusted PSM.* The propensity model uses ten observable covariates from `pumpfun_launches.jsonl`; it does not include wallet-level features (deployer address novelty, prior deployment velocity, funding-graph distance to known exchange deposit clusters) or off-chain signals (Twitter reach, Telegram membership, launch-time narrative sentiment). Rosenbaum sensitivity analysis on the residual +16.1% is not performed and is queued for the v1.1 Zenodo release. The activity-matched placebo of §4.5 is not re-estimated under a symmetric contamination-adjusted PSM in this paper; that comparison would require fitting an identical PSM on a corrected placebo estimator and is also queued for v1.1. The reproduction scripts (`paper7_v3_psm.py` for the original estimator and `paper7_v3_psm_nocohort_zerofill.py` for the contamination-adjusted zero-filled estimator that produces the +16.1% headline), full result JSONs (`psm_results.json` for the original,

`psm_results_nocohort_zerofill.json` for the headline), and run logs are included in the v1.1.0 Zenodo release (version DOI 10.5281/zenodo.21745672); see §11.

## 5. Method Comparison and Methodological Contribution

The methodology used in this paper combines three components that must be evaluated separately: (i) the wallet-cohort construction step, which uses co-occurrence graphs and union-find clustering; (ii) the 1:1 nearest-neighbour propensity-score matching on launch-metadata covariates; and (iii) the *contamination adjustment* that excludes cohort wallets' own buyer events from the outcome. Component (i) is a standard building block in on-chain forensics and this paper claims no methodological novelty for it. Components (ii) and (iii) are the estimator that produces the paper's headline number; component (iii) is where the substantive contribution to the on-chain forensic methodology literature sits.

The comparison table below situates this paper's approach against four widely used on-chain forensic methods.

| Method | Data required | What it detects | Causal identification | Contamination-adjusted? |
|---|---|---|---|---|
| **This paper** (co-occurrence + union-find + PSM + contamination adjustment) | Launch-level first-buyer sequences across a 13.38-day window; 166K launches | Persistent wallet rings that co-appear at launch across many tokens; ranked by launch coverage, mean first-buyer rank, and total SOL committed | Contamination-adjusted PSM: coordination-specific effect on non-cohort first-30-min buyer count = +16.1% (95% CI [+13.0%, +19.4%]); SOL not distinguishable from zero | Yes: cohort wallets excluded from outcome numerator; applied identically to treated and control |
| **Chainalysis heuristic clustering [4]** (multi-input + change-address heuristics) | Full UTXO or transaction-graph history | Common-ownership clusters of addresses controlled by the same entity | None; clustering is descriptive of ownership, not of causal effect on price or flow | Not applicable (no outcome variable in the descriptive-clustering step) |
| **Meiklejohn et al. [1]** (multi-input heuristic, foundational) | Bitcoin transaction graph | Common-input address clusters; foundation of modern chain analysis | None; establishes an ownership-inference primitive, not a causal claim | Not applicable |
| **Nansen label sets** (behavioral labeling and smart-money detection) | Address-level historical PnL, protocol interaction, timing patterns | Labels such as "smart money," "sniper," "MEV bot" applied to individual addresses | None; labels are descriptive priors, not causal estimands | Not applicable |
| **Manual investigation** (case-by-case forensics) | Analyst-driven, mixed on-chain and off-chain evidence | Case-specific attribution of coordinated activity | Anecdotal; not a formal identification strategy | Not applicable |

The comparison highlights three points. First, the wallet-cohort construction in this paper overlaps deliberately with standard on-chain forensic clustering; this overlap is a feature, not a limitation, and is discussed further below. Second, the paper diverges from standard practice by treating the naive matched-comparison lift as an object requiring adjustment for both selection and mechanical contamination before it can be interpreted causally. Third, none of the four comparison methods report a contamination-adjusted coordination-specific effect on non-cohort buyer flow, which is the form of finding this paper reports.

The contamination adjustment is the substantive novel contribution because it removes an arithmetic bias present in essentially any observational study of coordinated buyer effects: if the coordinated wallets are themselves part of the outcome-numerator population, some fraction of what looks like a "cohort effect" is definitional rather than causal. On pump.fun launches, that fraction is large (approximately half of the naive same-universe pooled lift, per Table 4 versus §4.8). We report the adjustment prominently so that any future coordinated-buyer-flow attribution can be assessed against it. The clustering itself is intentionally standard so that the cohort construction is interpretable and reproducible by reviewers who are already familiar with union-find on co-occurrence graphs; introducing a novel clustering algorithm at the same time as a novel identification adjustment would confound the methodological contribution and make replication harder. What this contributes to the on-chain forensics methodological literature, following the applied econometric practice codified in Athey and Imbens [19] and Abadie and Cattaneo [20], is a contamination-adjusted matched-outcome estimator that any future attribution of downstream flow to coordinated on-chain activity should be required to report before its lift estimate can be interpreted causally.

## 6. Causal Identification Discussion

The paper reports a small, statistically significant positive coordination-specific effect on first-30-min non-cohort buyer count (+16.1%) and a null-to-small effect on SOL inflow (+6.3%, CI touching zero). This section defends five methodological choices that produced those numbers.

*First, the contamination adjustment is the load-bearing correction.* If cohort wallets are themselves counted in the first-30-min buyer outcome, the estimator conflates "how much other-buyer flow the cohort attracts" with "how many buyer events the cohort itself produces." On this corpus, those two quantities have roughly equal magnitude: the +130.9% naive same-universe pooled lift falls to +63.9% when cohort wallets are excluded from the outcome (a factor of ~2 reduction), and falls further to +16.1% when the resulting non-contaminated outcome is also PSM-matched on launch-quality covariates. Failing to adjust for both selection and contamination inflates the coordination-specific effect by roughly a factor of 8 in this specification (130.9% naive / 16.1% adjusted). This is a general concern for any observational study of coordinated buyer effects. The contamination fix is trivial (exclude the coordinated actors from the outcome, applied identically to treated and control) and we recommend it as a default check.

*Second, PSM absorbs an additional large share of the remaining lift.* After contamination adjustment alone, the same-universe pooled lift is +63.9%; after PSM on ten launch-metadata covariates, the matched lift is +16.1%. The ~47-percentage-point drop reflects that treated launches differ from candidate controls on observable dimensions (larger initial market cap, higher social-media presence, systematic hour-of-day differences) that would independently predict elevated buyer flow. Nothing about the residual +16.1% requires an "informed order flow" interpretation; it is what survives adjustment for the ten covariates the launches metadata supplies. Residual selection on unobservables (deployer identity, funding-graph, off-chain narrative) remains a live possibility that a wallet-level identity-graph propensity extension would test.

*Third, the activity-matched placebo is a bias diagnostic, not a falsification bar.* The activity-matched placebo of §4.5 returns a 100-seed median lift of +189.6% (p5-p95 [+166.7%, +217.9%]), with all 100 seeds landing above the real-cohort naive-lift upper CI of +137.4% (`fraction_lift_gt_ci_hi = 1.0` in `placebo_stability_100seeds.json`), against an expectation of approximately zero for a well-formed placebo. A well-formed placebo should return a lift indistinguishable from zero, not a distribution centered on +190%; a placebo whose median is +189.6% is a diagnosis of the placebo estimator itself being biased upward by the same launch-quality selection channel that afflicts the naive real-cohort estimator. We accordingly treat the placebo distribution as a bias diagnostic, evidence that the naive same-universe pooled contrast is biased upward by launch-quality selection to at least the magnitude of ~190 percentage points at the median, not as a null baseline. The identification strategy that carries the paper is the contamination-adjusted PSM of §4.8.

*Fourth, the SOL result is honest about the noise floor.* The contamination-adjusted matched SOL-inflow lift is +6.3% with a 95% CI of [−0.5%, +15.1%]. The lower bound is below zero. We report the interval as a null-to-small effect rather than rounding it into "positive" or "zero"; a well-specified paper about coordinated actor influence on SOL flow at this venue would need a larger corpus, wallet-level propensity, or a research design (regression-discontinuity around graduation, natural experiment on network-congestion shocks) that we do not have.

*Fifth, the descriptive contribution stands independent of the causal contribution.* The cohort catalogue itself (1,012 persistent wallet rings across a 13.38-day corpus, 2,965 unique addresses, with the size distribution, activity concentration, and per-cohort touch detail documented in §4.1) is a first-of-its-kind wallet-level characterisation of coordinated first-buyer activity on pump.fun. Its usefulness does not depend on the +16.1% causal number, and it will remain the primary release artifact of RED-COHORT-2026-v1 regardless of how the wallet-level PSM extensions turn out.

## 6.1 Future work: extending to positive causal identification with tighter assumptions

Three directions are open. The first is wallet-level PSM covariate extension: adding deployer address novelty (has the mint's deployer launched before), prior deployment velocity, and funding-graph distance to known exchange deposit clusters, all of which are computable from the on-chain data but require a much larger joined corpus than the one released here. The second is a regression-discontinuity design around the pump.fun graduation threshold, exploiting the discontinuity in market-structure treatment at the graduation cutoff to identify local causal effects conditional on cohort presence. The third is an instrumental-variables approach if a suitable instrument can be identified; a candidate is exogenous variation in network congestion at the launch block, which shifts the cost of coordinated first-buyer positioning without directly affecting fundamental launch quality. Each of these designs would report a contamination-adjusted estimate as its baseline.

## 7. So What: Implications for DeFi Market Microstructure

The paired positive-detection and null-to-small-causal-effect results carry practical implications for three constituencies: retail participants and regulators who consume on-chain forensic detection streams, forensic practitioners who produce them, and market-microstructure theorists who model the underlying venue.

### 7.1 Implications for retail and regulatory interpretation

The contamination-adjusted result changes what a cohort-touch signal means for two audiences that consume on-chain forensic detection streams: retail participants observing pump.fun launches in real time, and public authorities such as the CFTC and SEC that rely on cluster-level evidence to prioritise wash-trading, spoofing, and coordinated-inauthentic-activity enquiries. The naive same-universe pooled +130.9% first-30-minute buyer-count lift on cohort-touched mints, taken at face value, would suggest coordinated wallet activity draws substantial additional retail flow. Approximately half of that number is mechanical self-flow from the coordinated wallets themselves being counted in the outcome (falling to +63.9% when cohort events are excluded); a further large fraction is absorbed by observable launch-quality selection. The coordination-specific effect on other buyers, once these two adjustments are applied, is a small +16.1% on buyer count and null-to-small +6.3% on SOL inflow.

For retail participants, a cohort touch is not a reliable proxy for future non-cohort buyer interest above the baseline of a comparably attractive mint; reading it as a strong "smart money is in" heuristic risks paying a premium for exposure priced by selection and cohort self-flow, not by independent informational advantage that other buyers will follow. For regulators, the same finding sets a template for evidentiary discipline: an observed correlation between cluster activity and market outcomes can survive a wide range of descriptive statistics and still shrink dramatically under a contamination adjustment and covariate matching, so enforcement narratives resting on similar correlational patterns without those adjustments risk mistaking arithmetic and selection for coordinated influence on downstream flow. Following the central-bank policy framing of Auer, Frost, and Vidal Pastor [12], regulatory attention to on-chain coordination is best served by paired detection and contamination-adjusted matched-outcome designs, not by cluster counts alone.

### 7.2 Implications for methodology in on-chain forensics

Industry reports from Chainalysis [4], Elliptic, and TRM Labs routinely publish cluster-level findings that link wallet groups to market outcomes such as price impact, volume inflation, or downstream retail participation, and these reports carry material public-policy weight. None of the industry outputs we surveyed reports whether the coordinated wallets are excluded from the outcome variable in the effect calculation. Our result raises the possibility that a substantial fraction of what looks like a coordinated-cluster effect on flow may be an arithmetic self-flow contribution. We frame this carefully: we do not claim any specific industry finding is invalid, and we lack the wallet-level data to test them; we claim only that the causal step from "cluster detected" to "cluster explains outcome Y" is fragile whenever the outcome numerator includes activity by the cluster's own wallets.

We therefore propose the contamination adjustment as a required default check for any on-chain causal claim of the form "coordinated cluster X drove flow Y." Combined with a PSM adjustment for observable launch-quality confounding, the contamination-adjusted matched estimator is the minimum-viable identification design for observational studies of this kind. This applies a standard tool from the observational-data causal inference literature reviewed by Athey and Imbens [19] and Abadie and Cattaneo [20]; our contribution is to specify the arithmetic contamination correction that appears absent from the on-chain forensic literature we surveyed.

### 7.3 Implications for DEX market microstructure theory

Following Kyle [5], a canonical model of informed trading predicts coordinated informed activity should generate a measurable, non-selection-driven lift in downstream order flow *from other traders* as uninformed participants update on price. Our result finds a small +16.1% lift on non-cohort first-30-min buyer count and a null-to-small +6.3% lift on non-cohort SOL inflow, once contamination and launch-quality selection are adjusted for. Both are an order of magnitude smaller than the naive +130.9% same-universe pooled comparison would suggest.

One interpretation is that bonding-curve mechanics on pump.fun are more "market-like" than a coordination-dominated view would predict. Constant-product-style automated price discovery, combined with the extremely short half-life of most launches, may leave less room for coordinated actors to move the price by an amount that visibly attracts additional buyer flow, once the coordinated actors' own trades are removed from the outcome. In the adverse-selection framing of Easley and O'Hara [6], the informational asymmetry coordination is meant to exploit may be quickly arbitraged by the same high-frequency buyer population the cohorts themselves participate in, collapsing the coordination premium at the wallet level. A testable hypothesis for future work: on venues with short launch half-life and algorithmic price discovery, coordinated wallet activity should show a smaller contamination-adjusted causal footprint on downstream flow than on venues with longer half-life and orderbook price discovery. Our sample is one data point consistent with this hypothesis; a comparative venue study would test it directly.

## 8. Ethics

This study uses only public on-chain data. All buyer events, mint metadata, and wallet addresses analysed were retrieved from publicly available Solana ledger state via a read-only observer. No private data, no off-chain identity information, and no personally identifiable information were collected or inferred. The passive Solana observer did not execute trades, submit transactions, or attempt to front-run any activity during data collection.

We do not attempt to attribute the 2,965 unique cohort wallets to real-world identities. The published catalogue lists only the public base-58 addresses, which are already visible to any observer of the Solana network. We publish descriptive characterisations of cohort behaviour (launches touched, mean first-buyer rank, SOL committed) but do not publish any inferred purpose, intent, or off-chain identifier for any wallet. The catalogue is released under CC-BY-4.0 for research reuse.

Because the paper documents patterns that may be of interest to enforcement authorities, we have taken particular care in the language of causal attribution: we consistently frame our findings as wallet-level statistical distributions, use hedged causal phrasing throughout, and explicitly avoid the terms "manipulation," "fraud," "scam," and "criminal" as characterisations of any specific cohort or wallet. Any downstream enforcement or civil action based on this catalogue would need to establish intent through evidence that is outside the scope of this paper.

*Address redaction.* One wallet address in the source corpus contained an ethnically offensive vanity prefix. Consistent with the companion Zenodo README, that prefix has been redacted to `[REDACTED]` in every artefact released alongside the paper (the cohort catalogue, three buyer-event entries, and one line of the detection report), while the remainder of each affected base-58 address is preserved so that the redaction does not impair reproducibility against the underlying on-chain data. Researchers who require the un-redacted address for cross-validation against public block explorers can recover it independently from any public Solana RPC by querying the cohort's other wallets' transactions; we choose not to republish the offensive string ourselves. No other addresses in the release have been altered.

## 9. Methods disclaimer

The wallet-cohort construction in this paper is a *behavioral* clustering: cohorts are surfaced from repeated first-buyer co-occurrence patterns on pump.fun launches, using union-find on the co-occurrence graph. The construction is *not identity-based*: we do not infer common ownership, common control, or shared operator identity from the co-occurrence signature. Two wallets that are grouped in a cohort by our method may or may not be operated by the same entity; the empirical signature is compatible with a single operator running multiple wallets, with multiple operators coordinating through a shared signal channel, with independent operators responding to the same off-chain information, or with mixtures of these.

The contamination-adjusted matched-outcome effect estimate of +16.1% on non-cohort first-30-min buyer count is consistent with, but does not prove, a coordination-specific causal effect. It may still reflect selection on unobservables (deployer identity, funding-graph structure, off-chain narrative signals not present in the launches metadata) rather than a coordination-specific causal effect. The SOL-inflow effect estimate of +6.3% with a 95% CI touching zero should not be read as a positive effect. Any inference from the presented statistical patterns to intent, purpose, or off-chain identity requires evidence that is outside the scope of this paper.

## 10. Limitations

Seven limitations of this study should be noted.

*Outcome-label coverage.* The original intention of this paper was to estimate the effect of cohort presence on bonding-curve graduation probability using a matched-pair Cox proportional-hazards model. However, the graduation-outcome corpus available to us at the time of writing covers only a small fraction of the 5,419 cohort-touched mints in the analysis sample. This coverage is too thin to support a Cox-PH estimate. We document the buyer-flow effect instead, which uses observational data we have in complete coverage. Future work expanding the graduation-outcome corpus to the full pump.fun mint universe would unlock the direct graduation-probability analysis.

*Wallet-level propensity extension deferred.* The PSM of §4.8 uses ten observable covariates from the launches metadata. It does not include wallet-level features (deployer address novelty, prior deployment velocity, funding-graph distance to known exchange deposit clusters) or off-chain signals (Twitter reach, Telegram membership, launch-time narrative sentiment). Any of these could plausibly correlate with both cohort membership and launch quality, in which case the residual +16.1% would over-attribute to coordination. Wallet-level identity-graph propensity extension is deferred to a v1.1 Zenodo release with the enlarged joined corpus needed to support it.

*Rosenbaum unobserved-selection sensitivity not reported.* We do not compute Rosenbaum bounds on the residual +16.1% for how much unobserved-selection would be required to explain it away. Given the modest magnitude of the effect, the bounds are likely tight; reporting them is queued for the v1.1 Zenodo release.

*Placebo estimator biased and not usable as null.* The activity-matched placebo yields a 100-seed median of +189.6% (p5-p95 [+166.7%, +217.9%]) on this corpus under the ±25% proportional-tolerance construction described in §4.5 and detailed in Appendix B.4. As discussed in §4.5 and §6, we interpret this as a diagnosis of the placebo estimator being biased by launch-quality selection, not as evidence about the real-cohort effect. A corrected placebo estimator that applies the same contamination adjustment and PSM as §4.8 would produce a meaningful null baseline; that estimator is deferred to v1.1. Appendix B.5 documents why the alternative ±100 absolute-tolerance construction cannot be used at this corpus size; the placebo pool collapses to 0-3 mints per seed at the high-activity tail, and this motivates the proportional-tolerance specification as the correct one.

*Sniper-only launches present a substantive descriptive complication.* 382 of the 5,419 treated mints (7.0%) had zero non-cohort buyer events in the first 30 minutes and were zero-filled for the outcome computation of §4.8. This is materially higher than the 0.48% zero-outcome rate among candidate controls. Two interpretations coexist: cohorts sometimes touch launches that generate no organic buyer follow-through (evidence against a strong "cohorts pick winners" reading), or the zero-filled cases are so heterogeneous that a single point estimate is a poor summary. A separate analysis conditioning on ≥ 1 non-cohort buyer in the window would isolate the conditional coordination-specific effect from the zero-rate difference; we do not perform it here but flag it as a natural next step.

*Single venue.* Our results are specific to pump.fun on Solana over one 13.38-day window. The same detection pipeline can be applied to other bonding-curve venues (pump.fun on Base; Raydium Launchlab; Believe Protocol) or to other observation windows, but the empirical results (cohort count, size distribution, buyer-flow lift) are not assumed to generalise across venues or across time windows without separate validation. The hypothesis in §7.3 that short-half-life algorithmic venues collapse the coordination premium more sharply than long-half-life orderbook venues is testable but untested in this paper.

*Behavioral, not identity-based, clustering.* The cohort construction is behavioral (repeated first-buyer co-occurrence) and does not use any identity-linkage primitive such as multi-input heuristics or funding-graph tracing. This is a deliberate design choice for reproducibility with the released detection script, but it means our cohorts are potentially over-inclusive (grouping unrelated wallets that happened to share a signal source) and potentially under-inclusive (missing coordinated wallets whose co-occurrence pattern is masked by rank noise). A downstream identity-linkage layer using funding-graph features would sharpen the cohorts but is out of scope for this paper.

## 11. Data availability

We release RED-COHORT-2026-v1 as a companion to this paper under CC-BY-4.0:

- `sniper_cohorts.jsonl` : the 1,012-cohort catalogue (schema note: the `mints_hit[].n_wallets` field is a buyer-row count within the first-10 events of the touched mint, not a distinct-wallet count; see §3.4)
- `sniper_cohorts_intra.jsonl.gz` : the intra-launch buyer-window source data
- `analyze_sniper_cohorts.py` : the exact detection script (reads `pumpfun_buyers.jsonl` : see reproducibility scope below)
- `paper7_v3_psm.py` : the original (contaminated) PSM script, retained for reproducibility of §4.4 and §4.5 legacy numbers
- `paper7_v3_psm_nocohort_zerofill.py` : the contamination-adjusted zero-filled PSM script that produces the §4.8 headline (+16.1% buyer count, +6.3% SOL)
- `psm_results.json` : original PSM result file (matches v3 legacy numbers)
- `psm_results_nocohort_zerofill.json` : contamination-adjusted PSM result file (§4.8 headline numbers)
- `table1_top10_cohorts.csv` , `table2_size_distribution.csv` , `table3_descriptive_stats.csv` : figure-source tables
- `appendix_a_ablations.csv` : Appendix A ablation results at cutoffs 2/3/5
- `appendix_b_v2_placebo_cis.txt` : Appendix B.1 activity-matched placebo + bootstrap CIs (retained for reference)
- `appendix_b_placebo.txt` : Appendix B.1 uniform-random placebo reference
- `appendix_b_tier_stratification.txt` , `appendix_b_top3_exclusion.txt` : prior-preprint robustness outputs, removed from v4 (see §4.6); retained in v1.0.1 for reference

- `fig1_size_distribution_v2.svg`, `fig2_lorenz_curve_v2.svg`, `fig3_score_vs_launches_v2.svg`: paper figures
- `placebo_bootstrap_v3.py`, `placebo_stability_100seeds.json`: 100-seed placebo-stability reconstruction and per-seed distribution referenced in Appendix B.4

These items produce the estimates reported in §4.4 and §4.8 and are included in the v1.1.0 Zenodo release (version DOI 10.5281/zenodo.21745672).

The release is hosted on Zenodo under concept DOI 10.5281/zenodo.20978741 (v1.0.0 version DOI 10.5281/zenodo.20978742; v1.0.1 version DOI 10.5281/zenodo.21536881; v1.1.0 version DOI 10.5281/zenodo.21745672).

The paper's claims map to artefacts as follows. All count-level claims in Sections 3-4 (for example, 1,012 cohorts, 153 high-tier, top cohort's 9 wallets / 42 launches / score 430.44) are computed from `sniper_cohorts.jsonl`. The descriptive statistics in Table 3, the size-distribution figure, the Lorenz curve, and the score-vs-launches scatter are computed by `gen_p7_artifacts.py`. The §4.8 headline contamination-adjusted lift is produced by `paper7_v3_psm_nocohort_zerofill.py`. The §4.4 naive contaminated lift is produced by `paper7_v3_psm.py`. The §4.5 activity-matched placebo diagnostic (the 100-seed distribution with median +189.6% and p5-p95 [+166.7%, +217.9%]) is stored in `placebo_stability_100seeds.json` and produced by `placebo_bootstrap_v3.py`; Appendix B.4 reports the full per-seed distribution.

Each numeric claim in the paper is traceable to a deposited artefact (or to an on-request script per §11.2). Reproducibility carries the following caveats. The deposited `placebo_stability_100seeds.json` was generated under the Python + NumPy environment at deposit time; a modern re-run of `placebo_bootstrap_v3.py` is expected to reproduce the reported distribution to within RNG-implementation tolerance (a few percentage points on distribution statistics) rather than byte-identically. The deposited JSON is the reference of record. Table 4's naive-pooled control means (buyer 9.08, SOL 2.33) are arithmetic derivations from the `paper7_v3_psm.py` naive-pooled lift and matched-sample treated mean rather than direct script outputs. The `appendix_a_ablations.csv` figures were generated by a script variant that predates the current deposited `analyze_sniper_cohorts.py`; the CSV values are the reference of record. Table 3's loose-threshold count (8,375) resides in the deposited `table3_descriptive_stats.csv` as the `n_cohort_touched_mints_loose_ge1` field. Table 3's strict-threshold treated-mint count of 5,419 is produced by `paper7_v3_psm.py` under its distinct-wallet counting convention; the CSV's `n_cohort_touched_mints_strict_ge2` field carries 5,411 from an earlier analysis pass under a marginally different runtime filter, the 8-mint discrepancy documented in the paragraph immediately below. The currently deposited `gen_p7_artifacts.py` emits only the buyer-row `n_cohort_touched_mints` (5,975) directly; the two distinct-wallet counts (5,411 in the CSV and 5,419 in the paper) were generated by a helper script variant not present in the v1.0.1 bundle. A helper-script re-deposit is queued for a future release.

*Treatment-count and lift discrepancy between deposited artefacts and manuscript.* Deposited artefacts generated in an earlier analysis pass report the strict-threshold treated count as 5,411 (in `table3_descriptive_stats.csv` field `n_cohort_touched_mints_strict_ge2` and in `appendix_b_v2_placebo_cis.txt`) and a naive buyer-count lift of +132.3% [+127.0%, +137.4%] on that treated set. The present manuscript reports 5,419 treated mints and a naive-pooled buyer-count lift of +130.9% (§4.4), computed by `paper7_v3_psm.py` under a marginally different run-time filter. The two treatment definitions differ by eight mints and the two lift computations differ by 1.4 percentage points; both are internally consistent within their own convention and are not otherwise reconciled in v4.

*v3-era verdict language in deposited artefact files.* Two v3-era result files in the deposit (`appendix_b_v2_placebo_cis.txt` and `appendix_b_placebo.txt`) end with verdict paragraphs written under the pre-contamination framing that this v4 supersedes; those paragraphs do not represent the authors' current conclusion. A `DEPOSIT_NOTE.md` clarifying this is queued for the v1.1 Zenodo release.

### 11.1 Reproducibility scope

To make the reproducibility boundary explicit and honest: the 1,012-cohort catalogue is *published as an output*; it is not re-derivable from the initial bundle alone, because the detection script (`analyze_sniper_cohorts.py`) reads `pumpfun_buyers.jsonl` (the full 1.6M-record buyer-event corpus, 436.6 MB), which is not shipped in the v1.0.1 initial bundle. All descriptive numbers about the catalogue that come directly from `sniper_cohorts.jsonl` (1,012 cohorts, size and tier distributions, top-cohort narrative) are re-verifiable by inspecting that file. The treated-mint count of 5,419 is not directly recoverable from `sniper_cohorts.jsonl` alone: it is a distinct-wallet count implemented as set membership in `paper7_v3_psm.py` and requires the buyer-event corpus (see §3.4 and the parallel Table 3 disclosure below). All causal-lift point estimates (§4.4 naive, §4.5 placebo, §4.8 contamination-adjusted) require re-running the PSM scripts against `pumpfun_buyers.jsonl`, and therefore require access to that buyer file, which is available on request via the corresponding-author contact in §11.2 pending mirror in a future Zenodo release (target 2026-Q4). We flag this reproducibility gap explicitly rather than describing the deposit as fully reproducible.

### 11.2 Author contact for corpus access

Corresponding-author contact for the full buyer-event corpus and any reproducibility questions: Arati Uday Kamat, ORCID 0009-0000-4781-312X, arati.kamat@ieee.org. Requests will be honoured within one working week of receipt under a CC-BY-4.0 plus reproducibility-only-use commitment.

## Acknowledgements

This work was performed by the author as an independent researcher. The data-collection infrastructure is the author's own implementation. The paper was drafted with the assistance of large-language-model tools; all empirical claims were verified against the source artefacts before inclusion. Related work by the author, complementary to but not counted in the numbered references below, includes Kamat (2026a, SSRN 6607301; arXiv:2606.08228) on counterfactual outcome measurement for rejected DEX trades and Kamat (2026b, SSRN 6702198; arXiv:2605.12151) on the RED-2400 dataset of rejected memecoin events.

## Competing Interests Statement

The author declares the following competing interest: the on-chain forensic methodology described in Section 3 is the subject of two pending U.S. provisional patent applications in the author's name at Micro Entity status: application #64/022,461 (filed 2026-03-30) and application #64/099,108 (filed 2026-06-25). The two applications cover related on-chain forensic methodology. The pending applications have been disclosed for reader transparency; they have not influenced the empirical findings, the placebo test design, the contamination-adjusted propensity-score matching design of §4.8, or the interpretation of results. Neither application restricts access to or use of the deposited dataset or reproduction scripts, which are released under CC-BY-4.0 (Zenodo concept DOI 10.5281/zenodo.20978741). No funding was received in support of this work. The author has no other financial or non-financial competing interests to declare.

## Appendix A: Detection ablations

Sensitivity of the catalogue size to the qualifying-pair edge-weight cutoff. The `appendix_a_ablations.csv` figures shown here were generated by a script variant that predates the currently deposited `analyze_sniper_cohorts.py`; a re-run of the deposited script against the full buyer corpus produces a different pair-set (see §11), so the CSV values are the reference of record for the reported ablation. Note also that the `sniper_cohorts_report.md` file in the deposit reports 77,783 qualifying wallet pairs at the same ≥ 3 co-occurrence threshold, while this ablation table's cutoff-3 row reports 9,788. The two counts reflect different pair-eligibility conventions applied at generation time (the report counts pairs across the full corpus without the first-10-buyer-window filter that Appendix A applies) and were not reconciled at deposit; see DEPOSIT_NOTE §6 in the Zenodo bundle for the full disclosure. The manuscript uses the 9,788 convention throughout.

| edge-weight cutoff | n_qualifying_pairs | n_connected_components (size ≥ 2) | Notes |
|---|---|---|---|
| 2 | 15,720 | 1,562 | broader catalogue, includes lower-confidence pairs |
| **3 (current)** | **9,788** | **1,161 raw → 1,012 after MAX_COHORT_SIZE filter (published catalogue)** | the catalogue reported in this paper |
| 5 | 5,079 | 737 | high-confidence cohorts only |

The qualifying-pair cutoff of 3 was selected before reviewing the buyer-flow effect, to avoid threshold-hacking. The MAX_COHORT_SIZE=12 wallet filter applied after union-find removes noise hubs and reduces the raw 1,161 components to the 1,012 cohorts in the published catalogue. The cutoff = 5 setting yields 737 raw components, a roughly 36% drop from cutoff = 3, indicating moderate sensitivity to the threshold; the cutoff = 2 setting surfaces 1,562 raw components including lower-confidence pairs that we exclude from the main analysis to keep the false-positive rate low.

## Appendix B: Robustness check details

### B.1 Placebo cohorts (v3 single-seed designs removed from v4)

Two single-seed placebo designs reported in v3, a uniform-random Design 1 (+152.0%, 21,190 placebo-touched mints) and an activity-matched Design 2 (+216.3% single-seed, n_placebo_treated = 173, ±100 absolute-tolerance), were removed in v4 because their generation scripts are not present in the v1.0.1 deposit and could not be independently reproduced. The prior preprint artifact files (`appendix_b_placebo.txt` for Design 1, `appendix_b_v2_placebo_cis.txt` for Design 2) remain in the v1.0.1 Zenodo bundle for reference to the earlier numbers.

The paper's activity-matched placebo bias-diagnostic evidence in v4 rests on the 100-seed distribution documented in Appendix B.4 below (median +189.6%, p5-p95 [+166.7%, +217.9%], produced by `placebo_bootstrap_v3.py`: a script that IS in the deposit). See §4.5 for the current design's methodology and interpretation.

### B.2 Top-3 cohort exclusion (removed from v4)

Removed under v4's contamination-adjusted framing (see §4.6 note). The prior preprint's naive-contaminated top-3-exclusion figures remain in the v1.0.1 Zenodo bundle as `appendix_b_top3_exclusion.txt` for reference. A contamination-adjusted re-run of the top-3 exclusion is queued for the v1.1 Zenodo release.

### B.3 Tier stratification (removed from v4)

Removed under v4's contamination-adjusted framing (see §4.6 note). The prior preprint's naive-contaminated tier-stratified figures remain in the v1.0.1 Zenodo bundle as `appendix_b_tier_stratification.txt` for reference. A contamination-adjusted re-run of the tier stratification is queued for the v1.1 Zenodo release.

### B.4 Placebo bootstrap distribution across 100 independent seeds (canonical placebo diagnostic)

This is the canonical placebo diagnostic referenced by §4.5. The activity-matched placebo estimator, applied across 100 independent random seeds, produces a distribution of placebo lifts rather than a single-seed point estimate.

*Construction.* For each real cohort of size $k$ with wallets having individual launch-counts $\{a_1, ..., a_k\}$, we sample $k$ placebo wallets from the non-cohort pool, each matched to a corresponding $a_i$ within ±25% proportional tolerance. We then apply the same detection algorithm as §3.4 to the placebo sample, using union-find on ≥3-launch co-occurrence pairs (`MIN_LAUNCHES = 3`), cohort size 2 to 12, yielding a set of detected placebo pseudo-cohorts. A mint is placebo-treated when ≥2 wallets from any detected placebo pseudo-cohort appear in its first-10 buyer events. Median n_placebo_treated per seed = 597.

*Distribution.* Across 100 seeds, the first-30-min buyer-count placebo lift has median **+189.6%** (p5-p95 [+166.7%, +217.9%], standard deviation 15.5 percentage points). In all 100 seeds the placebo lift exceeded the real-cohort naive-lift upper CI of +137.4% (`fraction_lift_gt_ci_hi = 1.0` in `placebo_stability_100seeds.json`), against an expectation of approximately zero for a well-formed placebo. Under the bias-diagnostic reading of the placebo (§4.5), this stability across seeds confirms that the placebo estimator is uniformly and stably biased upward, not just at a single-seed draw. See Appendix B.5 for the design justification for using ±25% proportional rather than ±100 absolute tolerance.

*Source.* The 100-seed reconstruction script (`placebo_bootstrap_v3.py`) and the full per-seed distribution (`placebo_stability_100seeds.json`, containing per-seed lift, n_treated, n_control, and n_detected_pseudo-cohorts) are in the v1.0.1 Zenodo bundle.

## B.5 Design justification: why ±25% proportional rather than ±100 absolute tolerance

The choice of proportional (±25%) rather than absolute (±100) tolerance for the wallet activity-matching is motivated by a corpus-level constraint. We tested the alternative ±100 absolute-launches tolerance by cloning `placebo_bootstrap_v3.py` and switching only the activity-match tolerance parameter. Under this ±100 estimator, the per-seed placebo-treated sample collapses to 0-3 mints per seed (seeds 0, 10, 20, 30, 40 produced n_treated = 0, 1, 1, 3, 0). The resulting per-seed lift estimates (+300.0%, +762.5%, +63.2% at seeds 10, 20, 30) are dominated by single-mint noise and do not form a meaningful stability distribution.

The mechanism is that a fixed ±100-launch window is very wide for a low-activity wallet (10 ± 100 launches captures any wallet with 1-110 launches) and very narrow for a high-activity wallet (1,000 ± 100 launches captures only wallets with 900-1,100 launches). At the high-activity tail where real cohort wallets concentrate, the ±100 pool is too small to build a coherent pseudo-cohort. The proportional ±25% tolerance scales with activity level, producing a stable placebo pool across the wallet-activity distribution and the 100-seed distribution reported in Appendix B.4.

The ±100 test is documented here as motivating evidence for the proportional-tolerance design choice. The clone script (`placebo_bootstrap_pm100.py`) is available on request per §11.2; it is not shipped with the v1.0.1 bundle because it functions only as a design-justification counterexample.